\documentclass[%
prb,
 amsmath,amssymb,
preprint,
superscriptaddress,
]{revtex4-1}

\usepackage[english]{babel}
\usepackage{graphicx}
\usepackage{dcolumn}
\usepackage{bm}

\usepackage[utf8]{inputenc}
\usepackage[T1]{fontenc}
\usepackage{mathptmx}
\usepackage{etoolbox}
\usepackage{diagbox}
\usepackage{color}

\newcommand{\rev}[1]{\textcolor{black}{#1}} 
\newcommand{\revSI}[1]{\textcolor{black}{#1}} 

\begin{document}

\title{Stability and superstructural ordering of alkali-triel-pnictide clathrates A$_8$T$_{27}$Pn$_{19}$}

\author{Frank T. Cerasoli}
\affiliation{Department of Chemistry, University of California, Davis, Davis, California 95616, USA}

\author{Xiaochen Jin}
\affiliation{Department of Chemistry, University of California, Davis, Davis, California 95616, USA}

\author{Genevieve Amobi}%
\affiliation{Department of Chemistry, Iowa State University, Ames, Iowa, IA 50011, USA}

\author{Kirill Kovnir}
\affiliation{Department of Chemistry, Iowa State University, Ames, Iowa, IA 50011, USA}
\affiliation{Ames National Laboratory, U.S. Department of Energy, Ames, IA 50011, USA}

\author{Davide Donadio}
 \email{ddonadio@ucdavis.edu}
\affiliation{Department of Chemistry, University of California, Davis, Davis, California 95616, USA}
\date{\today}

\begin{abstract}
Clathrates are a class of inclusion compounds that offer various useful and surprising phenomena, including superconductivity, thermoelectricity, and the potential for high-density ion storage.
Stability conditions within the Alkali-Triel-Pnictide A$_8$T$_{27}$Pn$_{19}$ family of unconventional clathrates are investigated with high-throughput density functional theory calculations, establishing trends in formation energy, structural and electronic properties, \rev{and identifying several new potential candidates.}
Electronic structure calculations and first-principles molecular dynamics simulations show that \rev{all alkali guests are fully ionized in the framework, and that the stability of electron-exact clathrates correlates with the ionization potential of the guest atoms, while their rattling behavior is governed primarily by the size of the guest ions relative to the cages.}
Targeted reactive synthesis from elemental precursors is attempted, resulting in \rev{four previously unreported ternary phases, two of which are structurally characterized}. However, the predicted clathrate phases are not obtained.
Further analysis reveals that the stability of ATPn clathrate compounds containing heavy elements, such as bismuth, depends strongly on relativistic spin-orbit coupling, which is often neglected in high-throughput studies that compute formation energies.
Finally, chemically induced superstructural ordering is described in relation to Wyckoff sites in the prototypical type-I clathrate unit cell.
\end{abstract}

\maketitle

\section{Introduction}

The current demand for energy presents an enormous challenge, predominantly compensated through the exploitation of unsustainable fossil fuels.
Materials design offers improved techniques for harvesting renewable resources and effectively storing energy.\cite{eisenberg_preface_2005,armaroli_future_2007,jain_commentary_2013}
Recently, clathrate compounds have gained attention as promising thermoelectrics, superconductors, and containers for high-density storage of ions or small molecules.\cite{PhysRevLett.74.1427,fukuoka_synthesis_2004,nolas_physics_2014,zhu_superconductivity_2021}
Additionally, inorganic clathrates have been shown to exhibit unusual phenomena, like order-disorder phase transitions\cite{brorsson_orderdisorder_2021} and chemically induced superstructural ordering.\cite{dubois_ordering_2005,owens-baird_chemically_2020}
After the early discovery of clathrate hydrates in the 1960s, inorganic clathrates were synthesized, as tetrahedrally bonded frameworks made of group-IV semiconductors (Si and Ge), hosting alkali metal inclusions.\cite{silicon_clathrate_1965,ramachandran_synthesis_1999}
More recently, unconventional clathrate compounds that do not contain Si or Ge have been synthesized from aliovalent species, but the rules governing favorable compositions remain unclear.\cite{johnsen_crystal_2007,liu_antimony-based_2009,he_synthesis_2012,wang_unconventional_2018,owens-baird_iiiv_2020,yox_unprecedented_2021, yox_new_2023,yox_organizing_2024}
Understanding the stability and formation properties of inorganic clathrates paves the way to the synthesis of new materials in this class with enhanced properties that could be exploited for renewable energy harvesting and storage and electronic applications.

In this work, we conduct a density functional theory (DFT) study to predict the stability and electronic properties of the electron-balanced A$_8$T$_{27}$Pn$_{19}$ type-I clathrate family (A=\{Na, K, Rb, Cs\}, T=\{Al, Ga, In\}, Pn=\{P, As, Sb, Bi\}).
The family consists of 48 unique compounds, three of which have been synthesized previously.\cite{owens-baird_iiiv_2020}
\rev{Within scalar-relativistic DFT, nineteen compositions not yet synthesized are predicted to lie below the convex hull. However, fully relativistic calculations indicate that spin-orbit coupling (SOC) may destabilize the Bi-containing candidates.} Conditions dominating stability and framework size are established.
Several enticing properties of clathrates stem from the rattling dynamics of the guest atoms.\cite{powell_clathrate_1948,sales_filled_1997} For example, the large anharmonic displacements of the rattlers enhance phonon scattering and contribute to lowering the thermal conductivity of clathrates.
This behavior makes clathrates promising phonon glass electron crystal (PGEC) systems for thermoelectric energy conversion.\cite{david_m_crc_2005,takabatake_phonon-glass_2014,wang_phonon_2018}
Here we analyze the rattler behavior with \textit{ab initio} molecular dynamics (MD) and establish a connection with the observed stability trends.
Finally, superstructural ordering is described in terms of transformations on the independent Wyckoff sites.

\subsection{Unconventional Clathrates}

The clathrate crystal structure consists of nanometer-sized polyhedral cages that may encapsulate guest atoms or small molecules, which are not covalently bonded to the framework.
Clathrate compounds come in several polyhedral cage arrangements and in a variety of chemical compositions.\cite{ammar_clathrate_2004,guloy_guest-free_2006,dolyniuk_clathrate_2016}
The first inorganic clathrates discovered in the 1960s were made of a silicon framework and Sodium guests: type-I Na$_8$Si$_{46}$ and type-II Na$_x$Si$_{136}$.\cite{silicon_clathrate_1965}
Conventional inorganic clathrates have frameworks composed of group-IV tetrel elements (Si, Ge, Sn) and account for the majority of known inorganic clathrates. 
Unconventional clathrate frameworks can contain late transition metals (Cu, Ni, Zn, Au, Cd), group-III triels (Al, Ga, In), and group-V pnictogens (P, As, Sb, Bi).\cite{wang_unconventional_2018}
Table \ref{fig:eb_clath_table} chronicles the few electroneutral, tetrel-free clathrates that have been synthesized to date. Depending on their stoichiometry, these compounds are usually intrinsic or doped semiconductors with a narrow band gap, very low thermal conductivity, and potential for high thermopower.  
This study uncovers stability conditions in group III-V clathrate compounds by adapting the crystal structure of Owens-Baird {\it et. al.}\cite{owens-baird_iiiv_2020} to other electron-precise compositions of triels and pnictides.

\begin{table*}[t]
\begin{center}
  \footnotesize
  \centering
  \begin{tabular}{ | c | c | c | c | c | c | }
    \hline
    Clathrate & Year & Author & Cell Size & a (\AA) & Volume (\AA$^3$) \\ \hline
    Cs$_8$Cd$_{18}$Sb$_{28}$ & 2009 & Liu {\it et. al.}\cite{liu_antimony-based_2009} & V$_0$ & 12.1916(15) & 1812.1(4) \\
    Cs$_8$Zn$_{18}$Sb$_{28}$ & 2009 & Liu {\it et. al.}\cite{liu_antimony-based_2009} & V$_0$ & 11.7054(16) & 1603.8(4) \\ 
    K$_8$Zn$_{18}$As$_{28}$ & 2012 & He {\it et. al.}\cite{he_synthesis_2012} & V$_0$ & 10.7173(6) & 1230.99(12) \\ 
    Rb$_8$Zn$_{18}$As$_{28}$ & 2012 & He {\it et. al.}\cite{he_synthesis_2012} & V$_0$ & 10.7413(6) & 1239.28(12) \\ 
    Cs$_8$Zn$_{18}$As$_{28}$ & 2012 & He {\it et. al.}\cite{he_synthesis_2012} & V$_0$ & 10.7731(8) & 1250.32(16) \\ 
    Cs$_8$Cd$_{18}$As$_{28}$ & 2012 & He {\it et. al.}\cite{he_synthesis_2012} & V$_0$ & 11.3281(5) & 1453.69(11) \\ 
    Cs$_8$Cd$_{18}$Sb$_{28}$ & 2020 & Owens-Baird et. al.\cite{owens-baird_chemically_2020} & 8V$_0$ & 24.3160(5) & 14377.3 \\ 
    Cs$_8$Zn$_{18}$Sb$_{28}$ & 2020 & Owens-Baird {\it et. al.}\cite{owens-baird_chemically_2020} & 27V$_0$ & 28.66$\times$49.64$\times$20.26 & 28823 \\ 
    Cs$_8$In$_{27}$Sb$_{19}$ & 2020 & Owens-Baird {\it et. al.}\cite{owens-baird_iiiv_2020} & 8V$_0$ & 24.4620(8) & 14637.8(14) \\ 
    Cs$_8$Ga$_{27}$Sb$_{19}$ & 2020 & Owens-Baird {\it et. al.}\cite{owens-baird_iiiv_2020} & 8V$_0$ & 22.9497(8) & 12087.4(13) \\ 
    Rb$_8$Ga$_{27}$Sb$_{19}$ & 2020 & Owens-Baird {\it et. al.}\cite{owens-baird_iiiv_2020} & 8V$_0$ & 22.89620(4) & 12003.01(6) \\ \hline 
  \end{tabular}
\end{center}
\caption{Synthesized tetrel-free, electron-precise type-I clathrates. Unit cell volume is listed with respect to the type-I clathrate prototype volume V$_0$.}
\label{fig:eb_clath_table}
\end{table*}

Applying the Zintl-Klemm concept to inorganic clathrates requires only consideration of the framework bonds.
Guest atoms can donate electrons to the framework sites, fulfilling octet configurations and remaining confined by non-covalent interactions.\cite{powell_clathrate_1948,wang_unconventional_2018,chen_review_2021}
Each framework site bonds tetrahedrally, which requires four valence electrons per site to satisfy the counting rule, totaling 184 electrons in the type-I unit framework that comprises 46 framework sites and 8 guest atoms. 
Enforcing electron-precise compositions provides a viable constraint on the stoichiometry and yields a semiconducting band structure,\cite{nainani_enhancing_2012} enabling application opportunities in electronics and energy conversion.
The electron-exact configuration requires that ionic inclusions donate valence electrons. \rev{Our calculations indeed suggest that all alkali guests do so, but the energetic cost of ionization, which grows with the ionization potential (IP) of the alkali metals, penalizes the stability of compounds containing guests with high IP.}

\begin{figure}
  \centering
  \includegraphics[width=.9\textwidth]{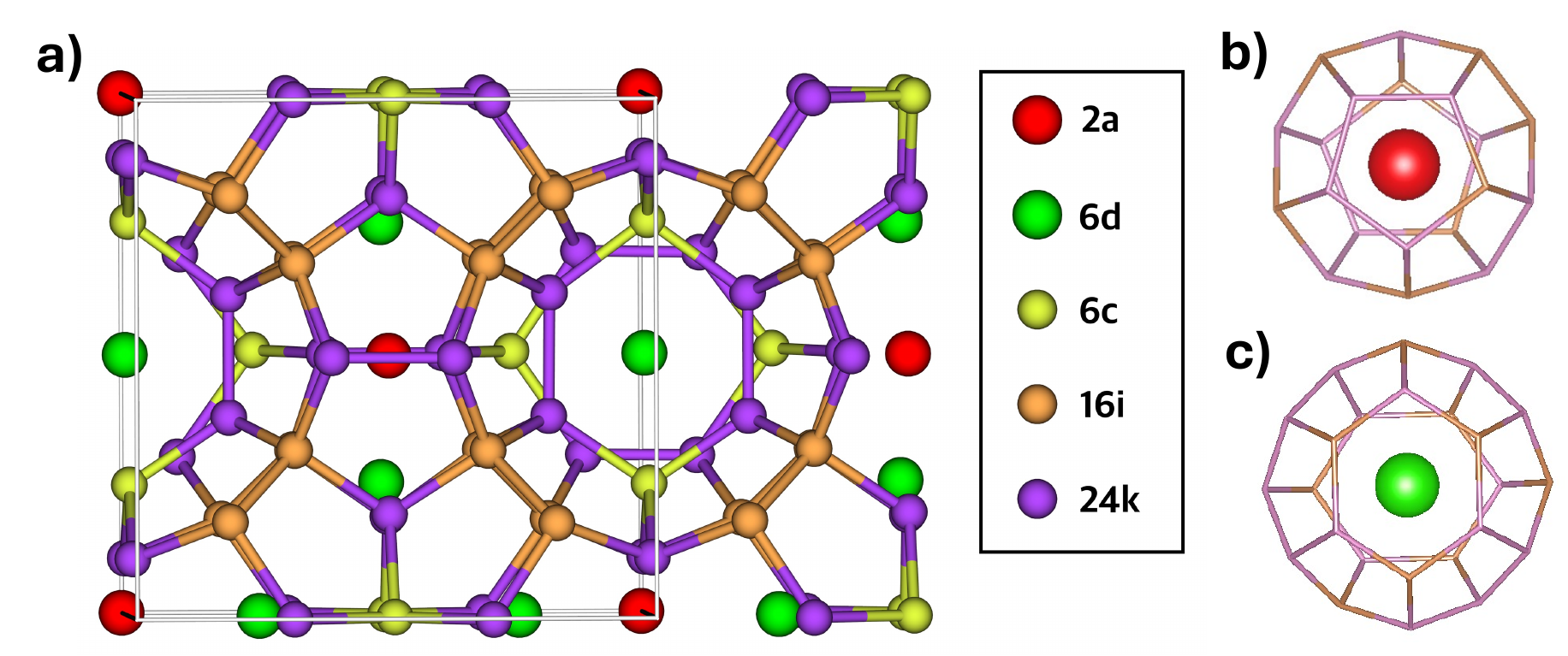}
  \caption{{(a) The type-I clathrate unit cell is shown within the cubic white boundary of side length $a$. 2a sites are shown in red, 6d in green, 6c in yellow, 16i in peach, and 24k in violet. Guest species are slightly enlarged compared to framework atoms. (b) The dodecahedron cage that encloses the 2a rattler is made of 12 pentagonal faces. (c) The tetrakaidecahedron cage that encloses the 6d rattler is made of 12 pentagonal faces and 2 hexagonal faces.}}
  \label{fig:clathrate_unit}
\end{figure}

Figure \ref{fig:clathrate_unit}a shows a type-I clathrate unit with lattice parameter $a$, which conforms to the cubic {\it Pm}$\bar{3}${\it n} space group.
Framework sites compose two types of polyhedral cages: regular dodecahedra with 20 vertices connecting 12 pentagonal faces (Figure \ref{fig:clathrate_unit}b), and tetrakaidecahedra with 24 vertices among 12 pentagonal faces and 2 hexagonal faces {(Figure \ref{fig:clathrate_unit}c)}. \cite{dolyniuk_clathrate_2016}
Five Wyckoff sites occupy the type-I unit, two (2a and 6d) for the 8 cage centers on which guest rattlers reside, and three (6c, 16i, and 24k) for the 46 framework positions.
Hexagonal faces are shared between adjacent tetrakaidecahedra, forming a cubic arrangement of cages, each enclosing one 6d site. 
2a sites are each enclosed by a pentagonal dodecahedron, which shares pentagonal faces with tetrakaidecahedra only.
Clathrate hydrates form with unit lattice parameters between 12.1~\AA\ and 12.2~\AA, owing to the slight differences in volume to inclusion type.\cite{takeya_crystal_2006}
Inorganic clathrates are generally denser and form in a wider range (10-12 \AA) of cell parameters, depending on framework composition rather than guest species.\cite{ramachandran_synthesis_1999,dubois_ordering_2005,allen_polyhedral_1964}
The structure is extensively cataloged, as the majority of known clathrates adopt this unit configuration.\cite{takeya_crystal_2006,guloy_guest-free_2006,johnsen_crystal_2007,christensen_clathrate_2009}

The A$_8$T$_{27}$Pn$_{19}$ clathrate family crystallizes in the body-centered {\it Ia}$\bar{3}$ space group, with a volume 8 times that of the {\it Pm}$\bar{3}${\it n} type-I unit cell prototype.\cite{owens-baird_iiiv_2020}
The superstructure is comprised of $8\times(27+19)=368$ framework sites and $8\times 8=64$ guest sites and features exact site occupancy.\cite{owens-baird_chemically_2020}
Superstructural ordering is common among unconventional clathrates, but, depending on synthesis protocols, crystals with mixed site occupancy may be obtained.\cite{yox_organizing_2024}
The type-I clathrates with composition A$_8$T$_{27}$Pn$_{19}$ are particularly interesting because they exhibit an ordered arrangement of triel and pnictogen atoms over the framework sites within a superstructure of dimension 2$a$ x 2$a$ x 2$a$. 
This 432-atom conventional superstructure may be reduced to a {\it bcc} unit cell containing 216 atoms. \rev{A Monte Carlo site-permutation search for alternative atomic configurations\cite{cerasoli_effective_2024} does not find arrangements significantly lower in energy than the experimentally observed ordering.}

\section{Methods}\label{sec11}

\subsection{Stability Calculations}
A Python workflow was developed as an engine for building electronic structure input files and initiating simulation environments automatically.
The engine selects the decomposition compounds and executes electronic structure calculations for all systems involved.
Testing this automated procedure on the three experimentally known A$_8$T$_{27}$Pn$_{19}$ clathrates (Cs$_8$In$_{27}$Sb$_{19}$, Cs$_8$Ga$_{27}$Sb$_{19}$, and Rb$_8$Ga$_{27}$Sb$_{19}$) reveals accurate determination of the clathrate energy with respect to the convex hull, when compared to previously reported theoretical predictions.\cite{owens-baird_iiiv_2020}
Ternary phase diagrams and crystal structures for the neighboring compounds were provided by the Materials Project Database, which was queried through the pymatgen\cite{ong_python_2013} library.
Self-consistent DFT calculations for stability evaluation were performed with the Quantum ESPRESSO\cite{giannozzi_quantum_2009,giannozzi_quantum_2020} open-source software library.
Scalar-relativistic pseudopotentials were generated with pslibrary v1.0.0,\cite{dal_corso_pseudopotentials_2014} using the projector augmented-plane wave (PAW) method.\cite{blochl_projector_1994}
The exchange-correlation functional was treated with the generalized gradient approximation (GGA), under the approximations of Perdew, Burke, and Ernzerhof (PBE).\cite{perdew_generalized_1996}
Plane waves are expanded on uniform Monkhorst-Pack (MP) k-meshes.\cite{monkhorst_special_1976}  
The enormous clathrate cells were computed at only the $\Gamma-$point, while MP grids for decomposition compounds are automatically chosen according to cell volume and number of atoms.
Kinetic energy cutoffs were taken as recommended by the PS-library, selecting the highest cutoff among elements in the clathrate composition.\cite{dal_corso_pseudopotentials_2014}
Decomposition compounds were computed with cutoffs consistent with the parent clathrate.
Every clathrate was optimized with respect to the atomic positions, enforcing that all forces fall below 0.04~eV/\AA.
Lattice parameters for each clathrate were chosen to minimize the Vinet equation of state.\cite{vinet_compressibility_1987}

\subsection{Molecular Dynamics}
\textit{Ab initio} MD simulations are performed with the CP2K Quickstep engine.\cite{kuhne_cp2k_2020}
NVT dynamics proceeded with 2 fs time steps at \rev{300~K and 600~K}, controlled with stochastic velocity rescaling.\cite{bussi_canonical_2007} 
For these simulations, we adopt the same GGA-PBE functional as in the stability calculation.
Valence electrons are expanded on molecularly optimized triple-zeta Gaussian basis functions,\cite{vandevondele_gaussian_2007} and core electrons and ions are treated by norm-conserving pseudopotentials in the Goedecker, Teter, and Hutter (GTH) form.\cite{goedecker_separable_1996,hartwigsen_relativistic_1998,krack_pseudopotentials_2005}
{In general, this setup gives results comparable to well-converged calculations for both molecular and covalent systems.\cite{miceli_liquid_2016,mirhosseini_uzh_2026} Here, we have verified that energy differences along MD trajectories are consistent within 10 meV/atom when computed with CP2K and \rev{Quantum ESPRESSO}.}
\rev{The 216-atom cells were simulated for 4~ps at 600~K and 2~ps at 300~K. 
Atomic displacements are characterized by the mean-squared displacement from the ideal clathrate configuration, MSD$(t)=\langle|\vec{r}_i(t)-\vec{r}_i^{\,\mathrm{id}}|^2\rangle$, where $\vec{r}_i^{\,\mathrm{id}}$ are the ideal positions, with the alkali guests at the cage centers. The MSD is averaged over all atoms of the same Wyckoff site (guests) or species (framework). For the guests, the MSD thus measures the mean-squared distance from the cage center.}
{Bader charges are computed a posteriori \rev{on MD frames}, postprocessing the charge density files obtained from CP2K runs.\cite{tang_grid-based_2009}
Bader} charges are averaged over all guests of the same unit Wyckoff site and across framework sites of the same species.

\subsection{Synthesis and Characterization}\label{sec:guided_synthesis}
Synthesis was attempted for some of the compositions with the most favorable stability predictions, with frameworks containing gallium or indium as the triel and bismuth or arsenic as the pnictide.
All reactants were weighed in an inert atmosphere of argon gas within a glovebox with O$_2< 1$~ppm due to the high reactivity of the alkali metals.
The synthesis routes included reactions of elemental precursors, pre-reacted binary compounds, and salt flux-assisted syntheses with a total reaction mass of 250 mg, excluding flux.
Various reaction containers, silica, alumina, and niobium, were utilized, coupled with diverse temperature profiles. See Supporting Information (SI) for the details of selected syntheses \revSI{ (Tables~S2--S10)}.

Initial characterization of the crystalline products was performed using a Rigaku Miniflex 600 diffractometer with Cu-K$\alpha$ radiation ($\lambda$ = 1.54185 \AA) at room temperature.
We used air-sensitive holders, which can be loaded in a glovebox, to avoid exposing the sample to ambient atmosphere.
Diffraction patterns were analyzed by comparison with known phases, and samples exhibiting novel or unidentifiable peaks were further examined using scanning electron microscopy with energy-dispersive X-ray spectroscopy (SEM-EDS).
Elemental analysis of samples was conducted using a JSM-IT200-SEM with an EDS detector (Dry SD25, JOEL, Inc., USA) and analyzed using the SMILE VIEW software.
Samples were mounted inside a glovebox on the homemade air-sensitive holder using double-sided carbon tape.
For structural characterization, single crystals were selected from reactions targeting Cs$_8$In$_{27}$As$_{19}$ and Rb$_8$In$_{27}$As$_{19}$.
Single-crystal X-ray diffraction (SCXRD) was conducted using a Bruker D8 Venture diffractometer equipped with a Photon100 CMOS detector and Mo-K$\alpha$ radiation ($\lambda$ = 0.71073 \AA).
Data collection was carried out at 173 K under a nitrogen stream to ensure the stability of air-sensitive crystals.

\section{Results}\label{sec2}
\subsection{Stability Prediction}\label{sec:stability}

\begin{figure}
  \centering
  \includegraphics[scale=0.5]{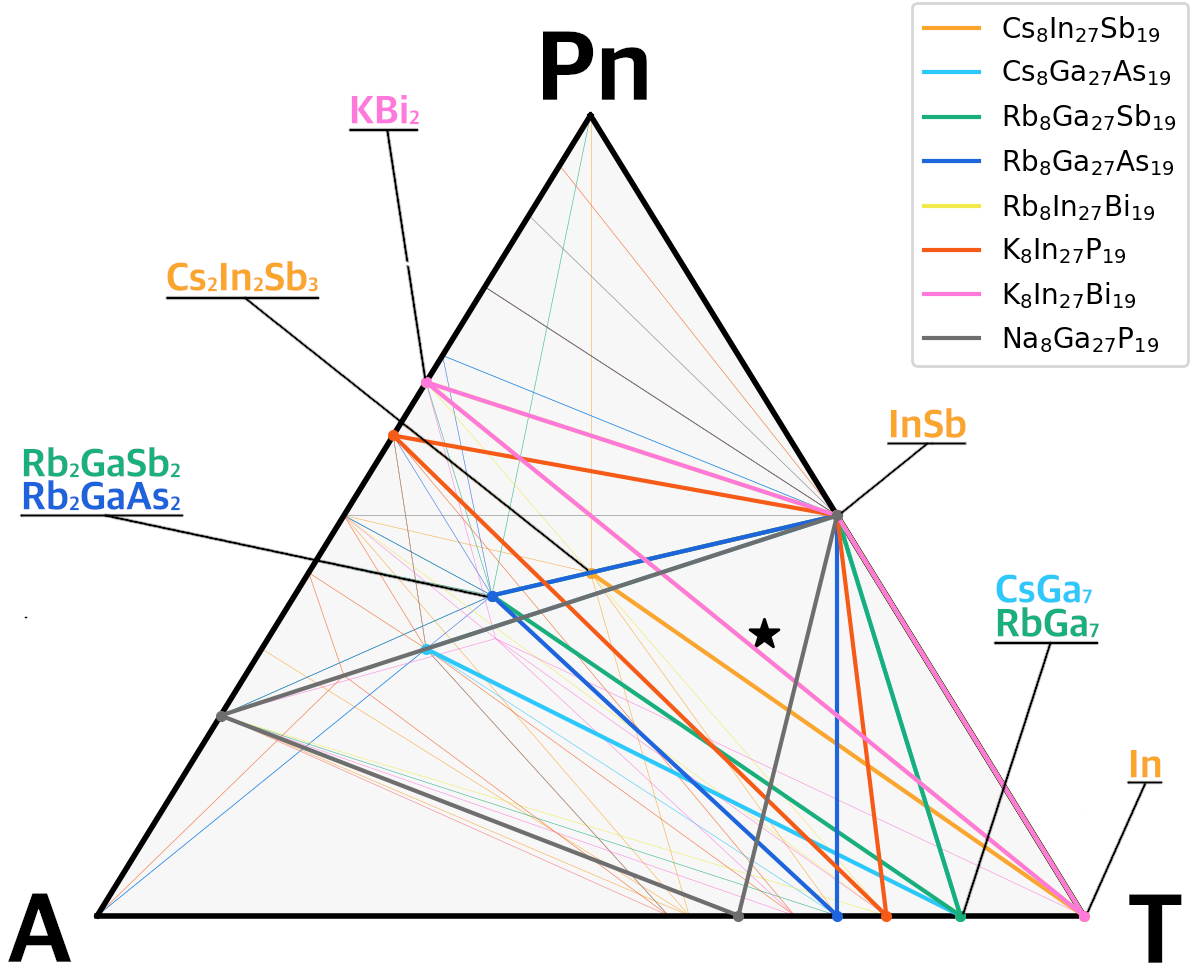}
  \caption{ Ternary compositional diagram for select compositions in the A$_8$T$_{27}$Pn$_{19}$ family, with a star representing the clathrate stoichiometry. Lines are overlaid to establish the convex hull for each respective composition, and facets enclosing the clathrate composition are embossed with thick lines. Select binary and ternary phases are labeled to demonstrate similarity between convex hulls, though the shape of each composition's hull varies dramatically.}
  \label{fig:phase_diagram}
\end{figure}

To predict stability, the formation energy of each clathrate compound is computed with respect to the convex hull of the compositions' ternary phase diagrams.
The clathrate location on the phase diagram, denoted as a star in Figure \ref{fig:phase_diagram}, determines which decomposition compounds an unfavorable clathrate structure would destabilize into.
Energies with respect to the convex hull are computed by comparing the clathrates to stable decomposition compounds provided through the {Materials Project database}.\cite{jain_commentary_2013}
Stability is calculated with the relation $E_{hull}=E_{clathrate}-\sum_iN_{i}E_{i}$, where $E_i$ are the decomposition compound energies and $N_i$ are the multiplicity coefficients required to balance the clathrate stoichiometry.
For example, the clathrate Cs$_8$In$_{27}$Sb$_{19}$ has neighboring compounds Cs$_2$In$_2$Sb$_3$, In$_1$Sb$_1$, and In$_1$ (elemental indium), which are labeled in Figure \ref{fig:phase_diagram}.
The energy of these four compounds is computed self-consistently and compared as:
\begin{equation}
E_{hull}=E_{Cs_8In_{27}Sb_{19}} - (4\cdot E_{Cs_2In_2Sb_3}+7\cdot E_{In_1Sb_1}+12\cdot E_{In_1})
\end{equation}

\begin{figure}[t]
 \centering
  \includegraphics[width=0.85\textwidth]{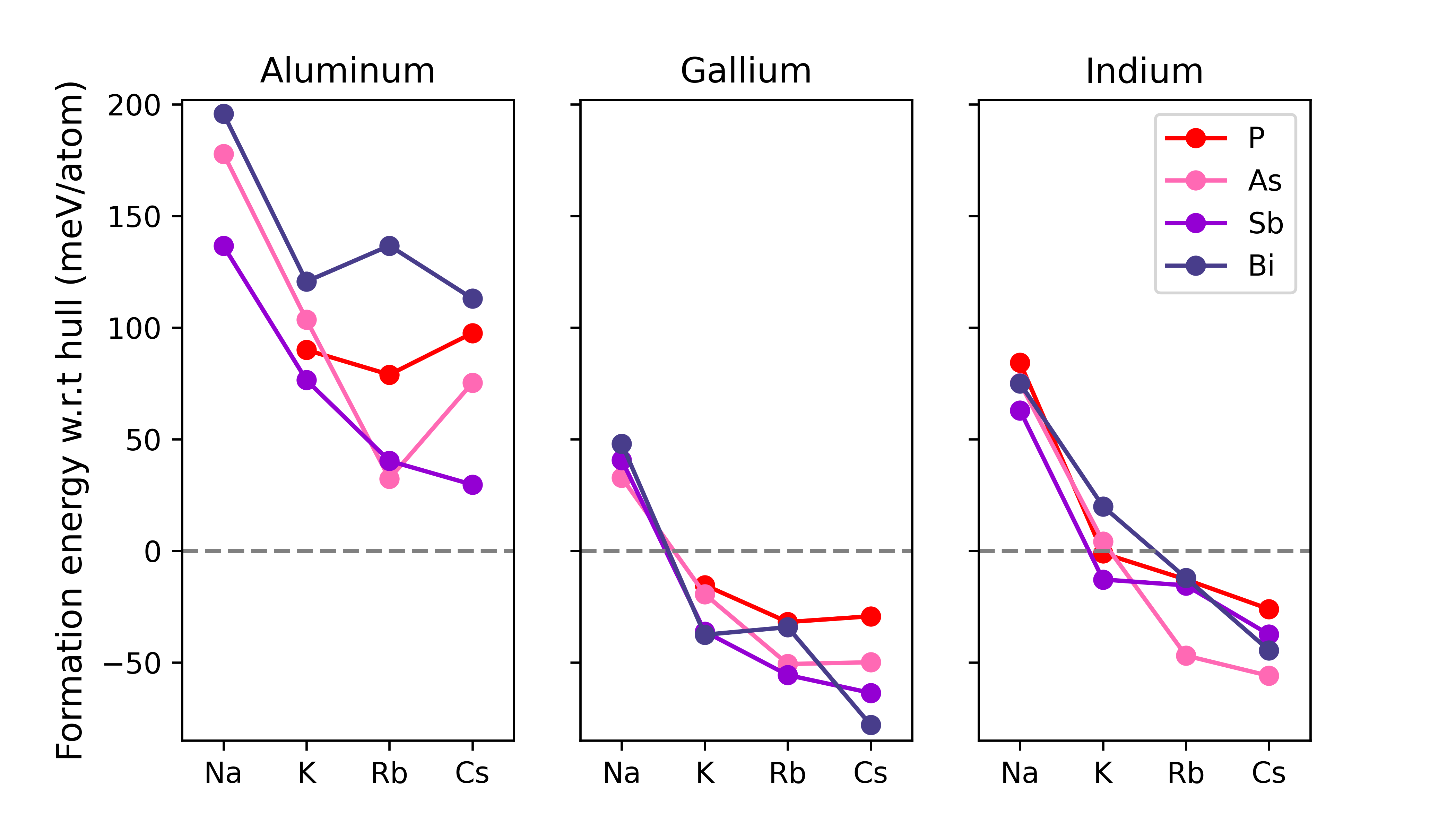}
  \caption{ Formation energy with respect to the convex hull is displayed as a function of inclusion species. Each triel is plotted in a different panel, and the pnictides are represented through color. Stability is shown to increase with the guest's atomic number. Lines are drawn between data points for clarity, and compounds containing phosphorus are omitted if framework-guest bonding occurred during optimization. \rev{Numerical values are listed in Table~S1 of the SI.}}
  \label{fig:guest_stability}
\end{figure}

The stability of A$_8$T$_{27}$Pn$_{19}$ clathrates in Figure~\ref{fig:guest_stability} is strongly affected by the guest composition and is most favorable when heavier elements occupy guest sites.
\rev{It is important to distinguish two possible origins of this trend. One would be incomplete charge transfer, \textit{i.e.}, light guests retaining their valence electron and leaving the framework electron-deficient. Our partial-charge analysis (Section~\ref{sec:lattice_dynamics}) rules this out: all alkali guests, including Na, are fully ionized, as Na carries a partial charge as large as that of Cs. The electron-precise framework is therefore realized for every guest, and the trend must instead originate from the energetic cost of producing the cations.}
Valence electrons in larger atoms are shielded by a greater number of core electrons and sit farther from the nucleus, lowering their ionization energy.
Sodium's ionization potential is markedly higher than that of the other alkali metals: within our computational framework, Na has an IP of 4.5 eV compared to 3.7, 3.5, and 3.2 eV for K, Rb, and Cs (PBE is known to underestimate absolute IPs, though relative differences are correct). Scaled to 8 alkali guests per formula unit, this difference in ionization cost translates to roughly 120 meV/atom between Na and K, and 190 meV/atom between Na and Cs.
\rev{These estimates exceed the stability differences in Figure~\ref{fig:guest_stability} (about 50--90~meV/atom between Na and K, and 80--140~meV/atom between Na and Cs for Ga and In frameworks) by roughly a factor of 1.5, which is expected because the free-atom IP neglects guest-dependent contributions to the lattice energy, such as the electrostatic stabilization of the cations in the cages and the strain of the framework. Hence, the IP  provides a consistent, semi-quantitative rationalization of the observed trend.}

\begin{figure}[hbt]
 \centering
  \includegraphics[width=17cm]{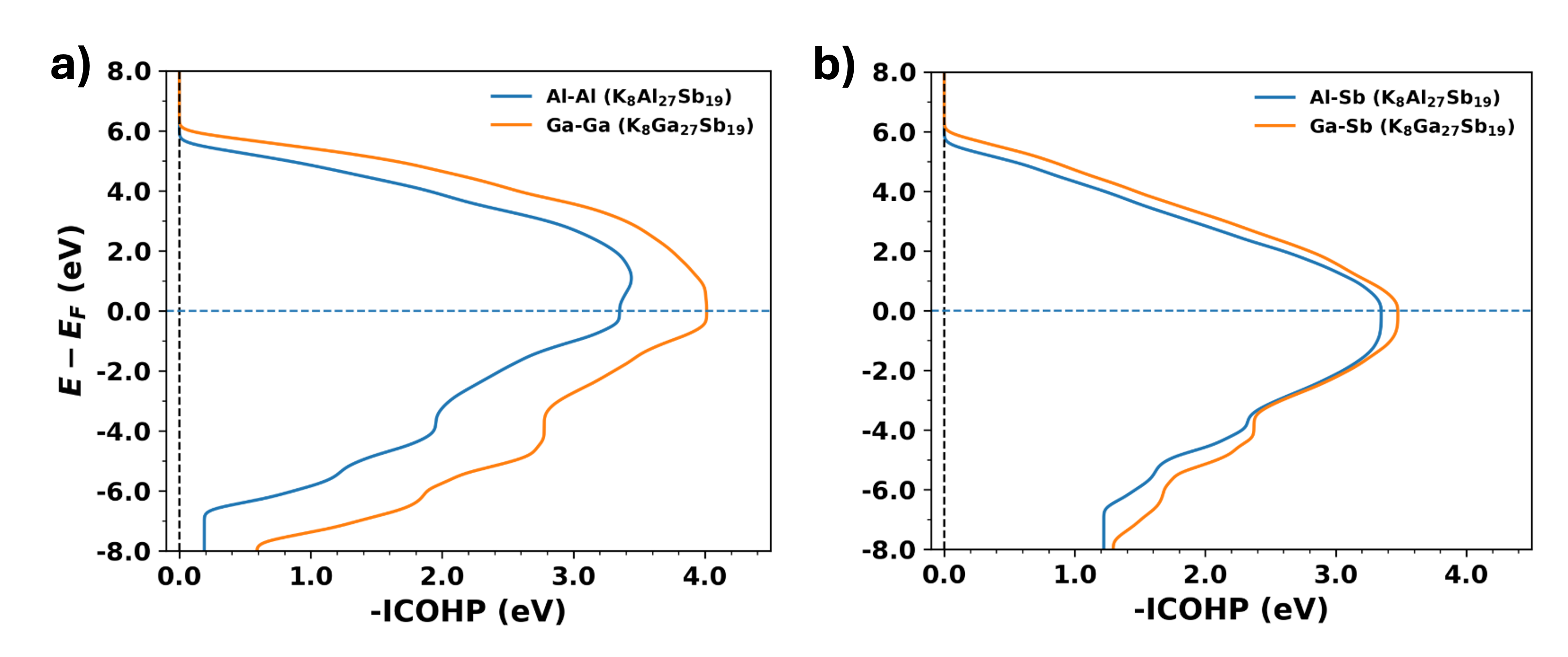}
  \caption{Negative Integrated Crystal Orbital Hamiltonian Population (-ICOHP) computed with Lobster\cite{maintz_lobster_2016} for homoatomic Al-Al and Ga-Ga bonds in K$_8$Al$_{27}$Sb$_{19}$ and K$_8$Ga$_{27}$Sb$_{19}$ (a), and for the heteroatomic Al-Sb and Ga-Sb bonds (b).}
  \label{fig:ICOHP}
\end{figure}
{Clathrates containing aluminum are always unstable, while compounds with gallium and indium are predicted to be stable when heavy guests occupy the framework.
The excess of triels in the A$_8$T$_{27}$Pn$_{19}$ stoichiometry makes inevitable T-T homoatomic bonds in the framework. 
We have calculated the crystal orbital Hamilton population (COHP)\cite{deringer_crystal_2011,maintz_lobster_2016} for both homoatomic and heteroatomic bonds in K$_8$Al$_{27}$Sb$_{19}$ (unstable) 
and K$_8$Ga$_{27}$Sb$_{19}$ (stable). Figure~\ref{fig:ICOHP} shows that the Ga-Sb and Al-Sb heteropolar bonds \rev{have very similar ICOHP values}, whereas the Ga-Ga bond exhibits a stronger bonding character than the Al-Al bond.}

Overall, \rev{within the scalar-relativistic calculations reported in Figure~\ref{fig:guest_stability}, 22} of the structures are predicted to be {energetically} stable, {19} of which currently remain unsynthesized.
\rev{Hull energies close to zero should not be read as a sharp classification. Within this work, the largest correction not included in the high-throughput workflow for Bi-free compounds is SOC, which shifts the hull energies of the As- and Sb-containing compounds by up to 10~meV/atom (Section~\ref{sec:SOC}).
Errors in the PBE reaction energies of ternary compounds relative to their binary constituents may be even larger, about 20--30~meV/atom,\cite{hautier_accuracy_2012} and vibrational free energies at synthesis temperatures are of comparable magnitude.
We therefore label as \textit{marginal} the compositions whose hull energy lies within $\pm10$~meV/atom of the hull, on either side, without changing the count of stable compounds given above.
With this criterion, K$_8$In$_{27}$P$_{19}$ (stable) and K$_8$In$_{27}$As$_{19}$ (unstable, +4.2~meV/atom) are marginal, and a further five stable compounds lie between $-10$ and $-15$~meV/atom (K$_8$In$_{27}$Sb$_{19}$, Rb$_8$In$_{27}$P$_{19}$, Rb$_8$In$_{27}$Sb$_{19}$, Rb$_8$In$_{27}$Bi$_{19}$, and Rb$_8$In$_{27}$As$_{19}$ once Rb$_2$In$_2$As$_3$ is included in the hull, see Section~\ref{sec:Rb2In2As3}).
The effect of SOC on this count is discussed in Section~\ref{sec:SOC}.}
{Further refinement of the stable candidates would require assessing their dynamical stability by computing their phonon dispersion. This is, however, too computationally costly to carry out at the level of DFT for 216-atom systems.}

\begin{figure}[hbt]
 \centering
  \includegraphics[width=12cm]{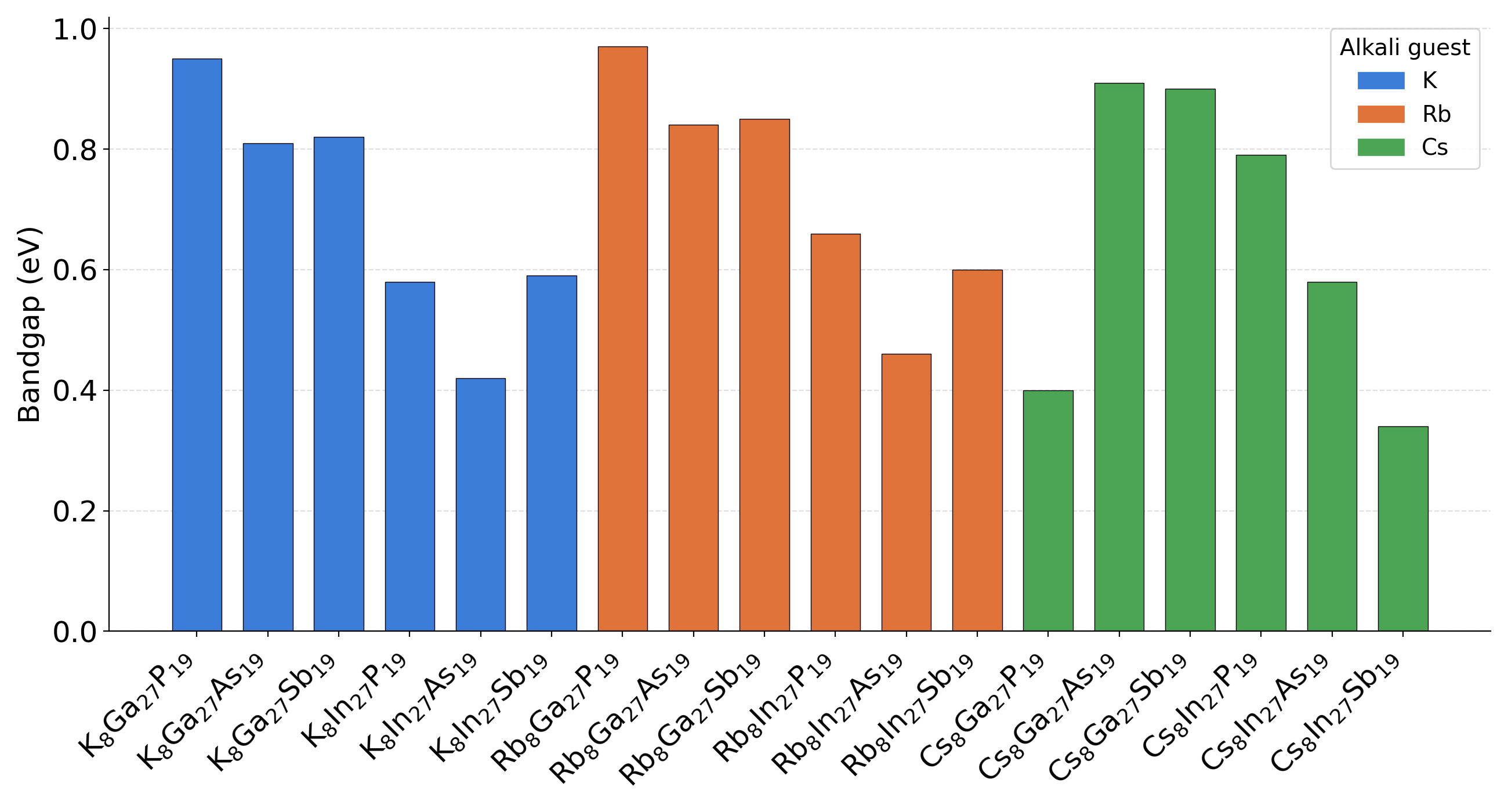}
  \caption{Kohn-Sham band gaps, computed at the $\Gamma$-point and sorted by guest alkali metal, of the \rev{Bi-free} clathrates predicted to be stable. \rev{K$_8$In$_{27}$As$_{19}$, which lies marginally above the hull (+4.2~meV/atom, Table~\ref{table:SOC}) and marginally below it when SOC is included, is shown for completeness. Bi-containing compounds are not included.}}
  \label{fig:bandgaps}
\end{figure}
{As expected from electron counting rules,\cite{owens-baird_iiiv_2020, yox_new_2023} all the compounds predicted to be stable are semiconductors. Their $\Gamma$-point band gaps, Figure~\ref{fig:bandgaps}, range from 0.34 eV (Cs$_8$In$_{27}$Sb$_{19}$) to \rev{0.97 eV (Rb$_8$Ga$_{27}$P$_{19}$)} and are mostly controlled by the combination of triels and pnictides in the framework when the guest atoms are K and Rb. The trend changes in Cs-containing clathrates, suggesting that Cs interacts with the electronic structure of the covalent framework in a non-trivial way, affecting the transport properties of the compounds. This was indeed the case for Cs$_8$In$_{27}$Sb$_{19}$ exhibiting extremely high hole mobility.\cite{owens-baird_iiiv_2020}}

\subsection{Synthesis Discussion}
To summarize our synthetic efforts, the major products of the reactions were known binary and ternary phases, as well as new ternary phases with various stoichiometries.
Most of the produced new ternary phases were highly air- and moisture-sensitive, which complicates detailed characterization of structure and composition.
Exploration within the A$_8$In$_{27}$As$_{19}$ phase space resulted in the discovery of two novel phases, Cs$_{2}$In$_{3.3}$As$_4$ and Rb$_2$In$_2$As$_3$, highlighting unreported competing ternary phases within this compositional region.
Similarly, for the K-Ga-Bi system, we have identified two distinct phases via SEM/EDX analysis with approximate compositions normalized to 1K as KGa$_{0.4}$Bi$_{1.7}$ and KGa$_{0.3}$Bi$_{0.7}$, despite the structure of those compounds remaining unknown.
These findings provide insight into the complexity of phase competition beyond the initially proposed clathrates. 
A comprehensive description of the synthetic procedures and reaction conditions is available in the {SI (Tables~S2--S10)}.
A summary of reaction products within the proposed ATPn system is presented in Table~\ref{table:synth_phases}.

\begin{table}
 \centering
  \begin{tabular}{c|c|c|c}
  \hline\hline
  Framework / Guest atoms & Cs  & Rb & K \\
  \hline
  In-As & Known phases +  & Known phases +  & Known phases + \\
   & {\textbf{Cs$_2$In$_{3.3}$As$_4$}} & {\textbf{Rb$_2$In$_{2}$As$_3$}} & *** \\
  \hline
    In-Bi & Known phases +  & Known phases +  & Known phases \\
    & *** & *** & \\
    \hline
      Ga-Bi & Known phases +  & Known phases +  & Known phases + \\
            &  *** & *** & {\textit{KGa$_{0.4(1)}$Bi$_{1.7(3)}$}} + \\
            &      &     & {\textit{KGa$_{0.33(8)}$Bi$_{0.7(7)}$}} + \\
            & & & *** \\
      \hline\hline
  \end{tabular}
  \caption{Reaction products obtained during synthetic trials. Newly discovered phases confirmed by single-crystal X-ray diffraction (SCXRD) are highlighted in bold, while those identified through energy-dispersive X-ray spectroscopy (EDS) are denoted in italics. Unidentified phases with distinct PXRD peaks are indicated with {***}.}
  \label{table:synth_phases}
\end{table}

\subsubsection{Crystal Structure of Rb$_2$In$_2$As$_3$}\label{sec:Rb2In2As3}
The compound Rb$_2$In$_2$As$_3$ was identified during exploratory synthesis targeting Rb$_8$In$_{27}$As$_{19}$. Isostructural to the A$_2$In$_2$Pn$_3$ compounds,\cite{ternary_antimonide_ownes_baird_2024,dipotassium_phyllotriantimonidodiindate_1991,dicaesium_phyllotriantimonidodiindate_1995} it adopts the Na$_2$Al$_2$Sb$_3$-type structure and crystallizes in the monoclinic \textit{P}2$_1$/\textit{c} space group. Rb$_2$In$_2$As$_3$ features two-dimensional (2D) layers composed of tetrahedral InAs$_4$ units connected by corner-sharing As atoms or covalent As–As bonds. All In atoms exhibit similar tetrahedral coordination environments, while the As atoms serve as corner-shared units linked to three InAs$_4$ tetrahedra or two InAs$_4$ tetrahedra and an additional As atom.
The bond lengths in Rb$_2$In$_2$As$_3$ (In–As) = 2.64–2.81 \AA; (As–As) = 2.50 \AA) align closely with those observed in the analogous K$_2$In$_2$As$_3$ compound. The compound was highly air- and moisture-sensitive, which prevented further characterization. 
The newly found compound is predicted to be stable within our DFT framework with an energy of $-87$ meV/atom with respect to the convex hull. 
Adding this newly found compound to the Rb-In-As compositional diagram, the  Rb$_8$In$_{27}$As$_{19}$ clathrate becomes significantly less stable, with the formation energy shifting from 47 to just 13 meV/atom below the convex hull.  

\begin{figure}
 \centering
  \includegraphics[scale=0.9]{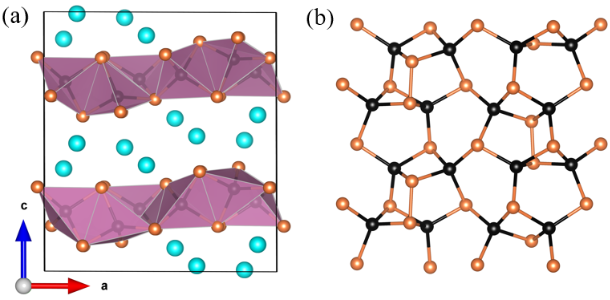}
  \caption{
  Crystal structure of Rb$_2$In$_2$As$_3$: (a) view along [010] 
  direction; (b) top view of the [In$_2$As$_{3}$]$^{–}$ layer. In atoms are
  shown in black, As atoms are shown in orange, and Rb atoms are shown in
  cyan. The unit cell is outlined in black.
  }
  \label{fig:synth_phases}
\end{figure}

\subsection{Stability Calculations with Fully Relativistic Spin-Orbit Coupling}\label{sec:SOC}

The unsuccessful synthesis of the bismuth-containing compounds predicted to be stable by DFT motivated further scrutiny of our calculations. For example, the formation energy with respect to the convex hull for K$_8$Ga$_{27}$Bi$_{19}$ was predicted to be -37.6 meV/atom, but the exact composition wasn't successfully synthesized. Therefore, we focused on bismuth-containing compounds, which we expected to be sensitive to corrections from considering fully relativistic spin-orbit coupling (SOC). 
Our refined calculations revealed that relativistic effects have a significant impact on the energetics of these compounds.
Bismuth is a heavy metal with 80 core electrons and a valence configuration of 6p$^3$.
The three unpaired valence electrons, in conjunction with strong core screening, cause spin-orbit effects that become crucial to bonding interactions.
It has long been understood that screening effects in heavy atoms result in larger, higher-energy outer shells. Yet, due to the increased computational cost, fully relativistic contributions are usually neglected in high-throughput calculations, including studies of stability or formation energy analysis of bismuth-containing compounds.\cite{pyykko_relativistic_1988}
In fact, SOC is responsible for the structural stability of the indium-bismuth binary In$_5$Bi$_3$.\cite{Chen_2020}

\begin{figure}
 \centering
  \includegraphics[scale=0.9]{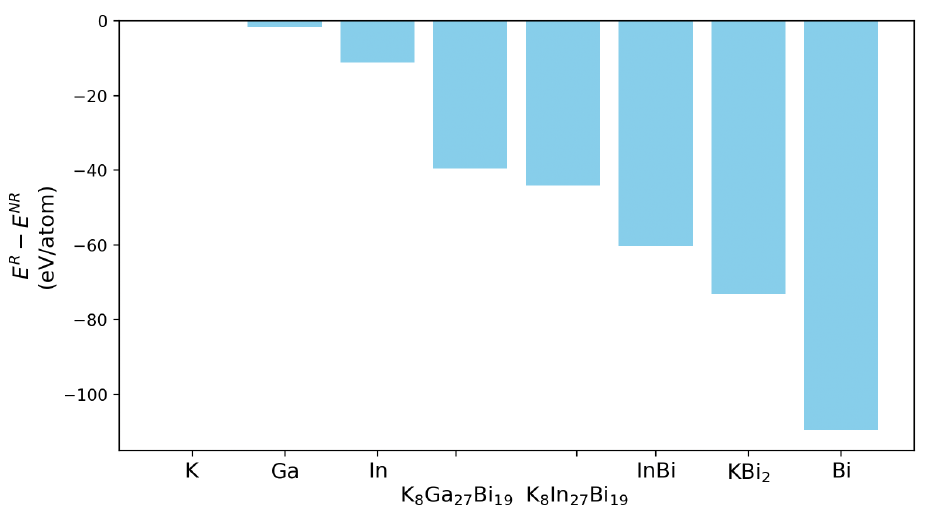}
  \caption{{Difference in DFT total energy before ($E^{NR}$) and after ($E^{R}$) incorporating fully relativistic spin-orbit coupling.}}
  \label{fig:soc_energy}
\end{figure}

In the case of A$_8$T$_{27}$Bi$_{19}$, neglecting SOC when using the scalar-relativistic pseudopotential leads to overestimating their total energies and stability.
As shown in Figure~\ref{fig:soc_energy}, the bismuth total energy undergoes a significant correction, while the potassium, gallium, and indium total energies are marginally affected by SOC. Meanwhile, considering fully relativistic SOC modifies the total energy of both the target clathrates and their competing phases. 
This leads to significant changes in the formation energy of Bi-containing clathrates on the respective convex hulls. The energies with respect to convex hulls of K$_8$In$_{27}$Bi$_{19}$ and K$_8$Ga$_{27}$Bi$_{19}$ increase by 30--40 meV/atom by incorporating SOC (Table ~\ref{table:SOC}). Remarkably,  K$_8$Ga$_{27}$Bi$_{19}$, which was predicted stable without incorporating SOC yet was not successfully synthesized, turns out unstable after the correction, with a formation energy increased to 5.2 meV/atom, leading to consistency with the experiment.
\rev{For clathrates containing lighter pnictogens, SOC shifts the energy relative to the hull by less than 10~meV/atom (K$_8$In$_{27}$As$_{19}$ and K$_8$In$_{27}$Sb$_{19}$ in Table~\ref{table:SOC}), in both cases in the stabilizing direction.
However small, this shift changes the sign of the hull energy of K$_8$In$_{27}$As$_{19}$, from +4.2 to $-$5.7~meV/atom. Since both values lie within $\pm$10~meV/atom of the hull, K$_8$In$_{27}$As$_{19}$ is classified as marginal in either treatment, consistent with the absence of the clathrate in our syntheses in the K--In--As system (Table~\ref{table:synth_phases}).
Fully relativistic calculations were carried out only for the four compositions in Table~\ref{table:SOC}. 
\rev{Among them, SOC moves K$_8$Ga$_{27}$Bi$_{19}$ above the hull and K$_8$In$_{27}$As$_{19}$ just below it, and it leaves the classification of K$_8$In$_{27}$Bi$_{19}$ (unstable) and K$_8$In$_{27}$Sb$_{19}$ (stable) unchanged. 
The number of compounds below the hull is thus unchanged when SOC is included for these four compositions, but both compounds that change sign end up in the marginal window (+5.2 and $-$5.7~meV/atom). 
For K$_8$Ga$_{27}$Bi$_{19}$, the SOC upward shift is large, about 40~meV/atom.}
If the destabilization by 30--40~meV/atom found for K$_8$In$_{27}$Bi$_{19}$ and K$_8$Ga$_{27}$Bi$_{19}$ transfers to the other Bi-containing clathrates, Rb$_8$In$_{27}$Bi$_{19}$ ($\approx -12$~meV/atom without SOC) would also become unstable, Rb$_8$Ga$_{27}$Bi$_{19}$ and Cs$_8$In$_{27}$Bi$_{19}$ would become marginal, and only Cs$_8$Ga$_{27}$Bi$_{19}$ ($-78$~meV/atom) may remain below the hull.}
In general, these results highlight the importance of incorporating SOC in high-throughput stability prediction of material systems that contain heavy elements such as lead and bismuth.

\begin{table*}[thb!]
\centering
\setlength{\tabcolsep}{4Mm}
\caption{Effect of fully relativistic spin-orbit coupling (SOC) on predicted formation energy $E_f$ (meV/atom) with respect to the convex hulls of \rev{selected K$_8$T$_{27}$Pn$_{19}$ clathrates. Negative differences indicate stabilization by SOC.}}
\label{table:SOC}
\begin{tabular}{cccc}
\hline
\hline
System & $E_f$ (original) & $E_f$ (with SOC) & Difference \\
\hline
K$_8$Ga$_{27}$Bi$_{19}$ & -37.6 & 5.2 & 42.8 \\
K$_8$In$_{27}$Bi$_{19}$ & 19.9 & 53.3 & 33.4 \\
{K$_8$In$_{27}$As$_{19}$} & {4.2} & {-5.7}  & {-9.9} \\
{K$_8$In$_{27}$Sb$_{19}$} & {-12.9} & {-16.8}	& {-3.9} \\
\hline
\hline
\end{tabular}
\end{table*}

\subsection{Lattice Constants}
Clathrate hydrate frameworks swell to accommodate larger guests,\cite{takeya_crystal_2006} and similar trends were revealed in clathrates with sp$^3$-hybridized framework bonds.\cite{guloy_guest-free_2006,he_synthesis_2012} 
{Conversely, the equilibrium volume of tetrel-free inorganic clathrates is {nearly independent of the guest alkali atoms and depends mostly on} the composition of the framework.}
The synthesized Cs$_8$In$_{27}$Sb$_{19}$, {Cs$_8$Ga$_{27}$Sb$_{19}$}, and Rb$_8$Ga$_{27}$Sb$_{19}$ compounds have measured lattice parameters 24.46~\AA, 22.95~\AA, and 22.90~\AA, which are in reasonable agreement with the respective computed lattice constants 25.11~\AA, 22.14~\AA, and 23.28~\AA, indicating that DFT-GGA is predictive for this property.

\begin{figure}
  \centering
  \includegraphics[scale=0.8]{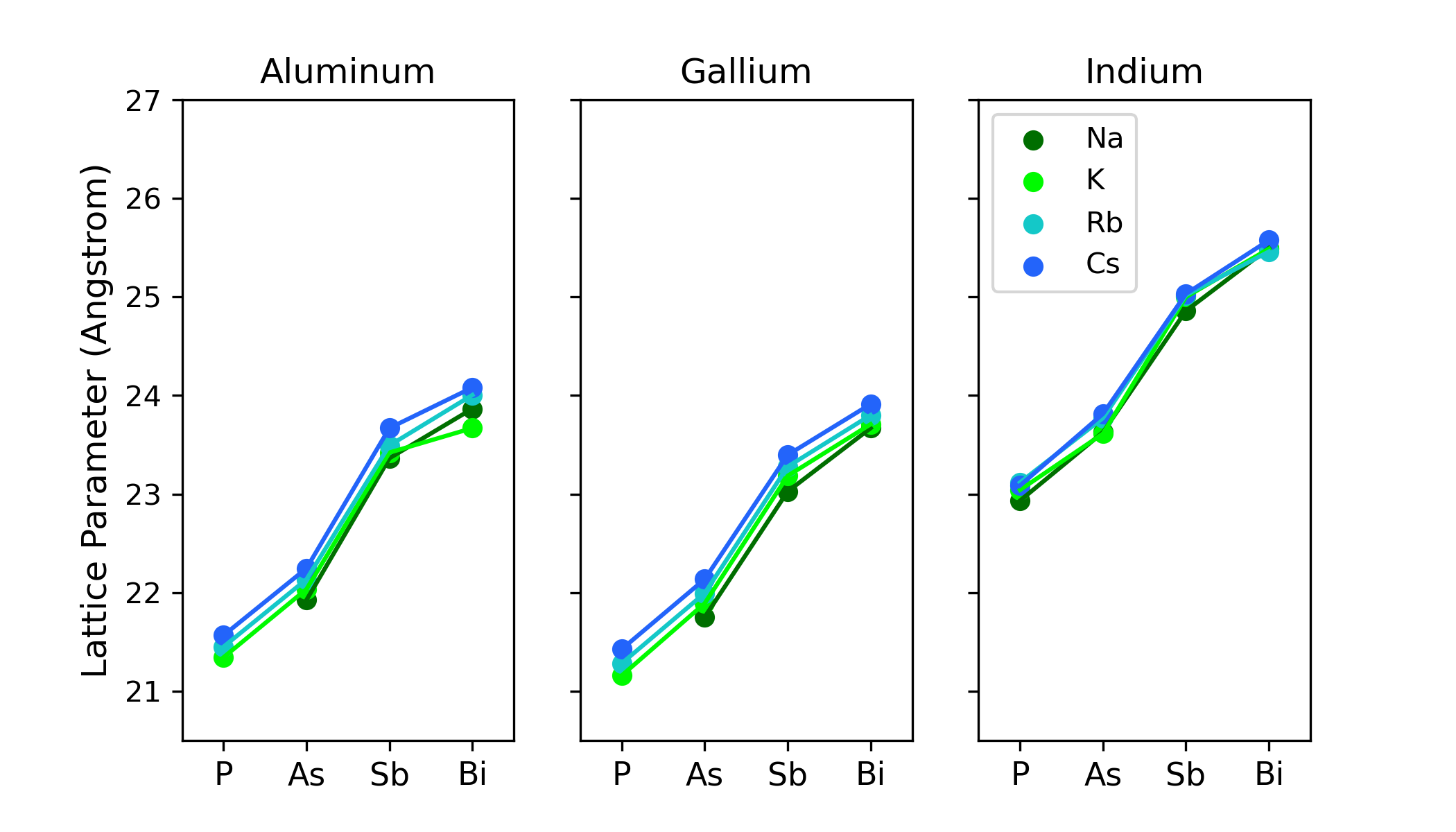}
  \caption{\footnotesize Lattice parameter in Angstrom, plotted against pnictogen type. Triels are each given an independent plot, and guests are represented through color. Framework size is shown to increase with the pnictogen's Z number. The framework size is predicted to be larger for indium-based compounds, again increasing with Z number.}
  \label{fig:lattice_constants}
\end{figure}

The lattice parameters for the 48 A$_8$T$_{27}$Pn$_{19}$ clathrates are reported in Figure \ref{fig:lattice_constants}.
In general, these calculations reveal that within the A$_8$T$_{27}$Pn$_{19}$ clathrate family, the cell size is primarily controlled by framework bonds, while the physical extent of guest atoms contributes very weakly. 
The Al and Ga series (left and central panels of Figure \ref{fig:lattice_constants}) have very similar lattice parameters, as the atomic radii of these two elements are similar. 
Volume dependence on framework species is supported by former studies on intermetallic clathrates, showing that the replacement of As with Sb increases the lattice parameter by 7.6\% in Cs$_8$Cd$_{18}$As$_{28}$ and by 8.7\% in Cs$_8$Zn$_{18}$As$_{28}$.\cite{he_synthesis_2012,liu_antimony-based_2009} The expansion is in agreement with the respective predicted increases of 5.4\% and 5.7\% for Cs$_8$In$_{27}$As$_{19}$ and Cs$_8$Ga$_{27}$As$_{19}$ upon complete substitution of As with Sb.
Refined stoichiometry in the synthesized arsenide compounds from previous works \cite{liu_antimony-based_2009,he_synthesis_2012} revealed only  80-90\% guest occupancy, while compounds containing antimony maintained full occupancy.
Vacant framework cages of inorganic clathrates are known to contract slightly,\cite{he_synthesis_2012} so this fractional difference in guest composition may account for the larger measured increase in lattice parameter.

\subsection{Dynamics of the Rattlers}
\label{sec:lattice_dynamics}

\begin{figure}
  \centering
  \includegraphics[scale=0.55]{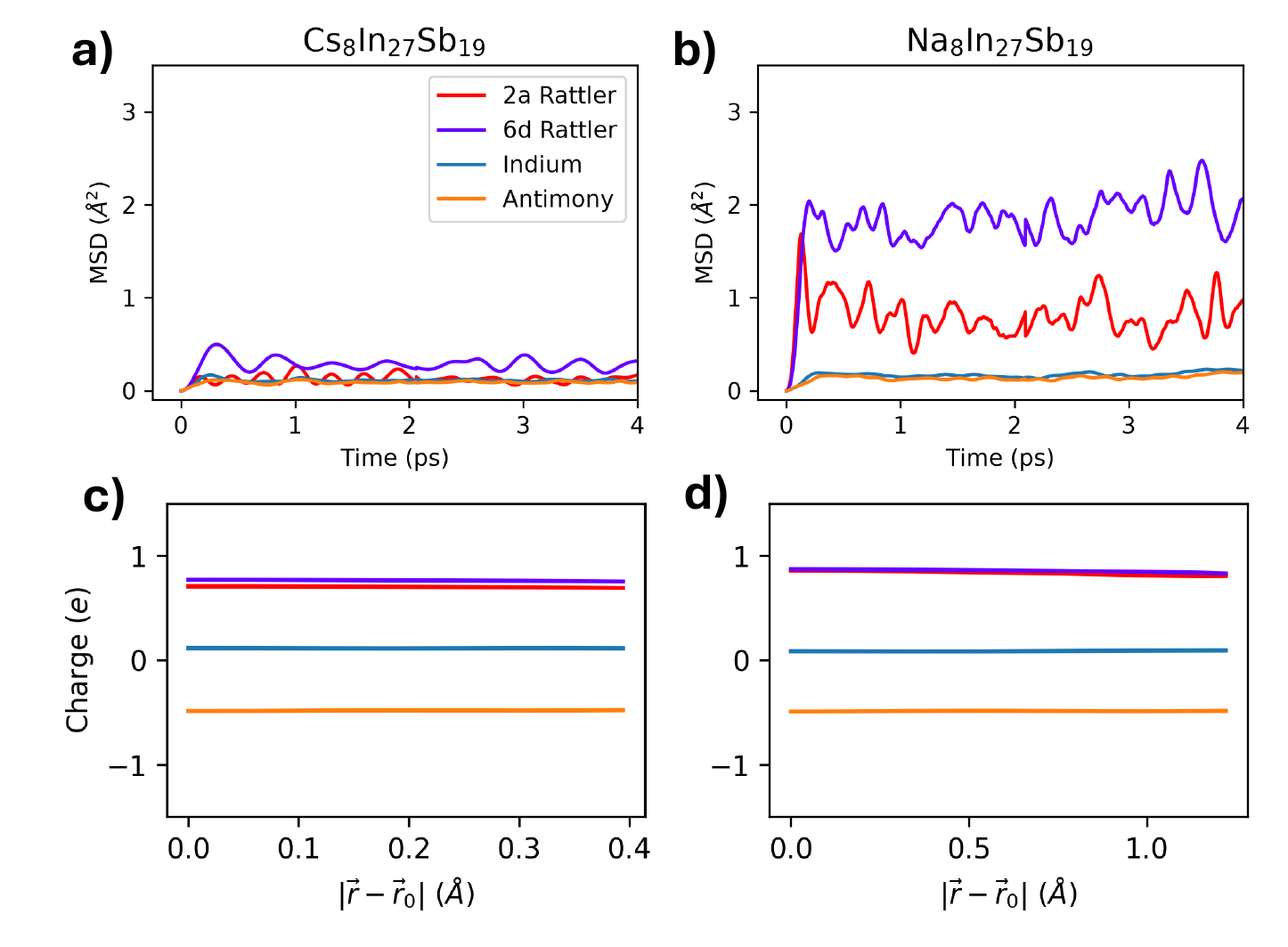}
  \caption{Molecular dynamics of Cs$_8$In$_{27}$Sb$_{19}$ (left) and Na$_8$In$_{27}$Sb$_{19}$ (right) across \rev{four} picoseconds at 600 K. \rev{The mean squared displacement from the ideal configuration with the guests at the cage centers, MSD$(t)=\langle|\vec{r}_i(t)-\vec{r}_i^{\,\mathrm{id}}|^2\rangle$, averaged over guests on the same Wyckoff site and over framework atoms of the same species,} is shown in panels (a) and (b),
  demonstrating that Cesium rattlers localize near cage centers, while Sodium rattlers tend to reside near cage walls. \rev{After an initial transient of about 0.2~ps, all MSDs fluctuate around stationary values.} \rev{Bader} charges are computed as a function of guest distance from the cage center $\vec{r}_0$ (panels (c) and (d)).
  $\vec{r}$ is the guest position, and the charges are averaged across similar sites and species. \rev{Both Cs and Na guests are fully ionized, independently of their position in the cage.} Cesium and Sodium achieve maximum displacements from the cage center of {0.39~\AA} and 1.22~\AA, respectively.}
  \label{fig:MD}
\end{figure}

First-principles MD simulations were performed to analyze charge localization on the guests and understand rattler behavior at finite temperature {with CP2K \cite{kuhne_cp2k_2020} (see Methods).} 
Sodium and Cesium, the guest species with the largest differences in size and ionization potential, were chosen to occupy an indium-antimonide framework.
\rev{Both clathrate frameworks remained intact over the simulated time at 300 and 600~K.}

As shown in Figure~\ref{fig:MD}a-b, the Cesium rattlers oscillate around the cage center due to thermal fluctuations\rev{; in the 300~K trajectories, their} root mean square displacement (RMSD) \rev{is} 0.26~\AA\ in the 2a position, and 0.41~\AA\ in the 6d position.
Conversely, Sodium rattlers are displaced from their crystallographic position and reside closer to the cage walls.
At 300~K, the Sodium atoms are displaced by 0.88~\AA\ from the 2a position and by 1.35~\AA\ from the 6d position.
6d guests incur larger displacements than 2a guests, \rev{since the tetrakaidecahedral cages are larger than the dodecahedral ones} (volume difference $\sim$ 40 \%).
\rev{The qualitative difference between the two guests is established within the first few hundred femtoseconds and persists over the whole trajectory: the MSD of the Cs guests remains bounded and small, whereas the Na guests leave the cage centers and explore off-center positions.
After this transient, all MSDs fluctuate around stationary values. At 600~K, the MSD of the Cs guests is about 0.15~\AA$^2$ on the 2a sites and 0.3~\AA$^2$ on the 6d sites, whereas that of the Na guests is about 0.8 and 1.8~\AA$^2$, respectively. 
The framework MSDs remain small in both compounds, about 0.1~\AA$^2$ in Cs$_8$In$_{27}$Sb$_{19}$ and 0.15--0.2~\AA$^2$ in Na$_8$In$_{27}$Sb$_{19}$: the off-centering of Na slightly enhances the vibrations of the framework but does not disrupt it over the simulated time.} 

{
An analysis of the charge density partitioning along the MD trajectories (\rev{Bader charges in} Figure~\ref{fig:MD}(c,d)) shows that both Na and Cs are completely ionized, and their partial charge is independent of their position within the cage.
\rev{The Na guests carry a partial charge (about $+0.85\,e$) at least as large as that of the Cs guests (about $+0.7\,e$), so the instability of Na-containing clathrates cannot be ascribed to incomplete electron donation to the framework.}
This finding makes it reasonable to directly correlate the formation energy of the clathrates \rev{with} the ionization potential of the guests, for each given composition of the framework.
Additionally, this observation suggests that the \rev{rattling behavior of the guests is governed primarily by their ionic radius relative to the cage size. Na$^+$, with an ionic radius} of 0.98~\AA, is subject to a flat potential at the center of the cage and is free to move toward the walls, whereas Cs$^+$, with a radius of 1.69~\AA, is restrained to the center of the cages.
}

\subsection{Superstructural Ordering}\label{sec:superstructure}

In this Section, we rationalize the atomic arrangement that leads to superstructural ordering in this class of clathrates.
Our hypothesis is that ordering within the 2$a$ x 2$a$ x 2$a$ cell is motivated by the maximization of heterogeneous bonding.
Reduction of the homogeneous bond count is achieved through symmetry operations on each independent Wyckoff site.
The number of Pn-Pn and T-T bonds is reduced by 24 when the atomic coordination is appropriately transformed from the base unit structure, producing 48 additional T-Pn bonds in the entire superstructure.
Only 16 Pn-Pn bonds exist within the superstructure, made possible through an intricate and specific ordering.

\begin{figure}
  \centering
  \includegraphics[scale=0.6]{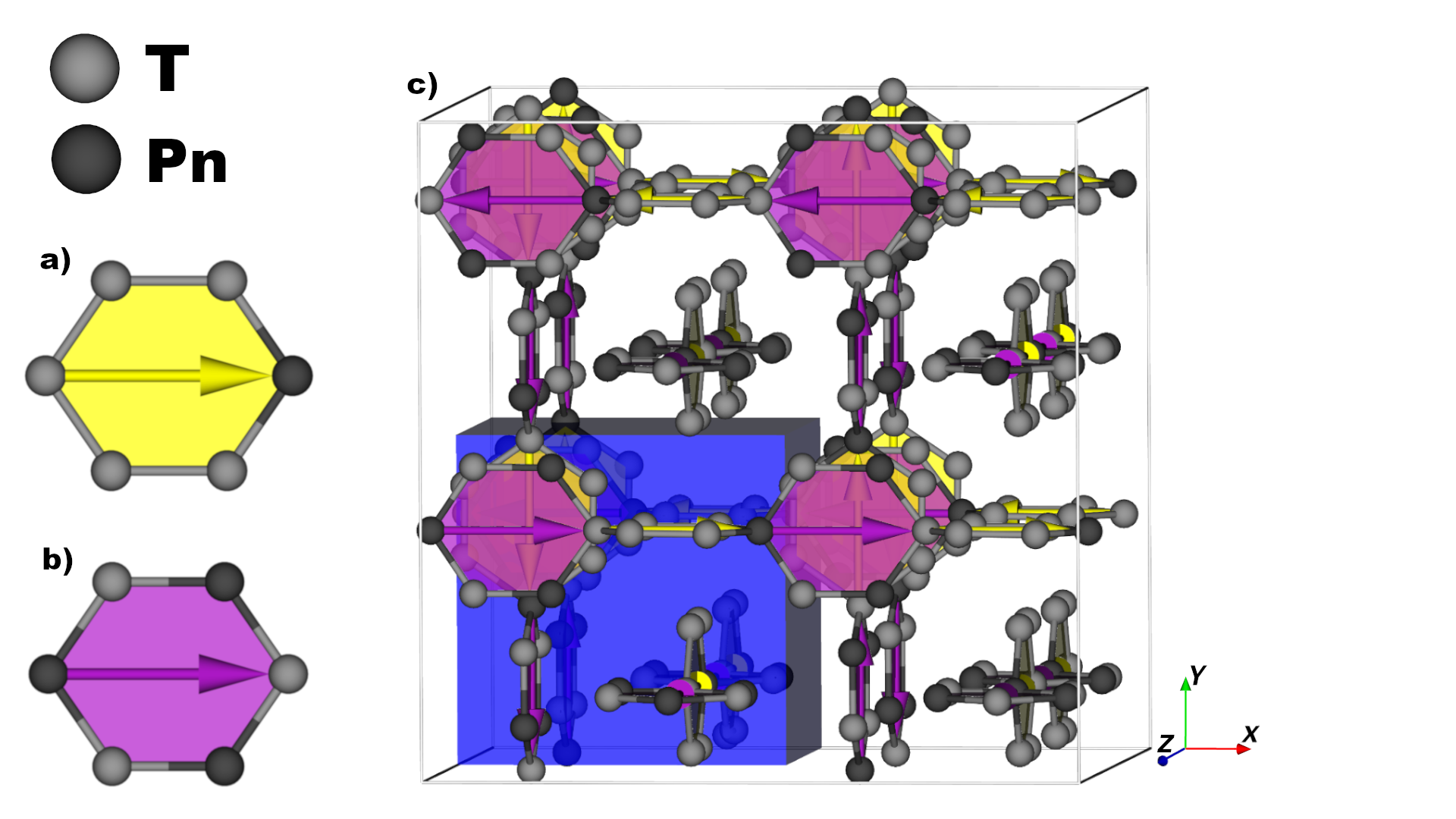}
  \caption{\footnotesize Chain network formed by the 6c and 24k sites, with 16i and guest sites omitted. The 2$a$ x 2$a$ x 2$a$ superstructure features chains of alternating direction in each Cartesian orientation. Panels (a) and (b) define the two unique arrangements of 6c+24k sites and their orientations. The faces are colored to identify relative composition, and arrows are displayed to clarify the direction of each hexagonal link. The superstructure exhibits alternating chain direction for adjacent chains of the same orientation (panel (c)), while the 46-atom unit framework (shaded in blue) can host only one chain per Cartesian orientation.}
  \label{fig:chains}
\end{figure}

The superstructure is now discussed in terms of the prototype type-I Wyckoff sites and unit lattice parameter $a$.
The 6c and 24k sites compose three orthogonally oriented chains of hexagonal faces, woven throughout the crystal along the Cartesian directions $\hat{e}_x$, $\hat{e}_y$, and $\hat{e}_z$.
The hexagonal planes form two types of chain link, shown in panels (a) and (b) of {Figure~\ref{fig:chains}}.
6c sites along the chains always alternate between the T and Pn species, while the 24k sites are ordered in alternating groups of four.
25\% of the 24k sites are occupied by pnictides and 75\% are occupied by triels.
The 2a x 2a x 2a supercell doubles the number of chains in each direction, illustrated in Figure \ref{fig:chains}c, and features reversed direction in adjacent chains.
Assigning coordinate $\vec{c}$ to an arbitrary 6c or 24k atomic site, sites forming adjacent chains are related by \textit{klassengliche} transformations.
The positional coordinate parallel to the chain direction $\hat{e}_i$ is related between sites on adjacent parallel chains by the transformation $\vec{c}_{i}\rightarrow-\vec{c}_{i}+\frac{a}{2}$, while the other two cartesian directions are related by a simple translation of length $a$ along the respective directions.
Hence, an atom with coordinate $\vec{c}_u=\langle c_{ux},c_{uy},c_{uz}\rangle$ along a chain with direction $\hat{e}_y$ corresponds to constituent positions $\vec{c}_{v1}=\langle c_{ux}+a,-c_{uy}+\frac{a}{2},c_{uz}\rangle$ and $\vec{c}_{v2}=\langle c_{ux},-c_{uy}+\frac{a}{2},c_{uz}+a\rangle$ on the two unique adjacent chains.\\

\begin{figure}
  \centering
  \includegraphics[width=.49\textwidth]{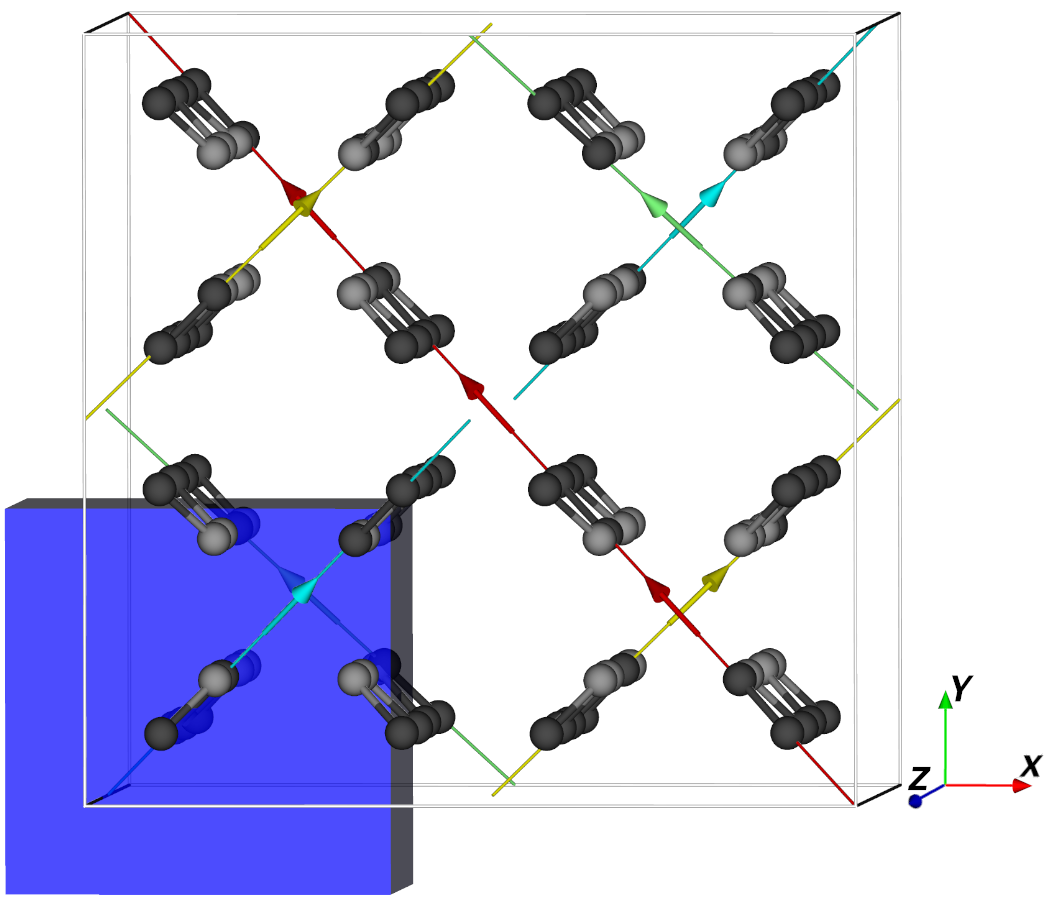}
  \caption{\footnotesize Pn-Pn bonds align linearly throughout the crystal, also collinear with 2a guest sites. Atoms are shown for 16i sites only, with arrows placed on the 2a sites in the direction of the aligned pnictides. Four colored tubes distinguish which cube diagonal the Pn-Pn bonds occupy. Again, T (Pn) atoms are displayed in light (dark) gray. The white boundary is shifted by $\langle \frac{a}{4},\frac{a}{4},-\frac{a}{4}\rangle$ with respect to Figure \ref{fig:chains}, however, the blue unit boundary highlights the same region in both figures.}
  \label{fig:capillaries}
\end{figure}

The 16i sites also contribute to the superstructural ordering, but under a different set of symmetry operations.
Pairs of pnictides form in a collinear fashion between 2a sites and adjoin the dodecahedral cages.
16i sites are arranged on the eight corners of cubes centered on each 2a guest site, partially composing the surrounding cage.
Three eighths of the 16i sites are occupied by triels, with the remaining sites occupied by pnictides.
The 16i cube corners surrounding one of the 2a guests host eight 16i pnictides, while 16i sites surrounding the other 2a guest are occupied by two pnictides and six triels.
The dodecahedral cages, which encapsulate the 2a guest sites, are connected to each other by a single bond between two 16i sites, 75\% of these bonds being T-Pn and 25\% Pn-Pn.
Figure \ref{fig:capillaries} illustrates the colinear Pn-Pn structure, with four unique orientations (one for each unique main diagonal of a cube) within the 2$a$ x 2$a$ x 2$a$ superstructure.
16i sites are related to each other by a mirror transformation dependent on the direction of cell replication. 
A translation of unit distance $a$ along $\hat{e}_x$ mirrors the surrounding 16i sites across a plane normal to $\hat{e}_y$ and centered at the 2a guest position enclosed in the blue boundary of Figure \ref{fig:capillaries}.
Displacement along $\hat{e}_y$ ($\hat{e}_z$) mirrors surrounding 16i sites across a plane normal to $\hat{e}_z$ ($\hat{e}_x$) and centered at the same 2a site.
This results in four unique orientations of 16i sites, each appearing twice in the 2$a$ x 2$a$ x 2$a$ superstructure.

Stable superstructures are formed as an assembly of units with \textit{klassengliche} transformations performed uniquely to each Wyckoff site, which arrange to maximize the heterogeneous bond count.
In the case of A$_8$T$_{27}$Pn$_{19}$ clathrates, the cell replication results in a superstructure with 8 times the prototype unit volume. 
This family of inorganic clathrates demonstrates that chemical composition is capable of lowering symmetries and minimizing pnictide bonding strictly through atomic decoration.

\rev{To test the sensitivity of our results to the assumed ordering, we performed Monte Carlo site-permutation searches (MC-SPS),\cite{cerasoli_effective_2024} with energies evaluated by a MEGNet graph-network model.\cite{chen_graph_2019}
For Rb$_8$Ga$_{27}$Sb$_{19}$, one of the experimentally known compounds, we have previously shown that MEGNet ranks different decorations consistently with DFT and that MC-SPS with simulated annealing, starting from random decorations on the DFT-relaxed geometry, recovers exactly the experimentally observed ordering in 29 out of 50 trials.\cite{cerasoli_effective_2024}
On the ideal type-I geometry, the same searches also identified an alternative $Ia\bar{3}$ decoration with the same count of homoatomic and heteroatomic bonds, which differs from the observed one by the exchange of the two species in half of the heteroatomic pairs on the 6c sites and in 25\% of those on the 24k sites. DFT places it within 2~meV/atom of the observed ordering.\cite{cerasoli_effective_2024}
Here, we consider a marginal (K$_8$In$_{27}$Sb$_{19}$) and an unstable compound Na$_8$In$_{27}$As$_{19}$.
After ten independent searches from random decorations for each compound \revSI{(Tables~S11 and S12 and Fig.~S1 of the SI)}, none of them recovered the observed ordering, and all the final arrangements are slightly higher in MEGNet energy, by 0.5--2.6~meV/atom for K$_8$In$_{27}$Sb$_{19}$, and 2.9--9.8~meV/atom for Na$_8$In$_{27}$As$_{19}$, and contain more homopolar bonds. 
For Na$_8$In$_{27}$As$_{19}$, the final arrangements contain only one to three additional In--In plus As--As pairs per cell (0.7--2.1\% fewer heteroatomic bonds), while for K$_8$In$_{27}$Sb$_{19}$ they contain six or eight (4.2--5.6\% fewer).
For K$_8$In$_{27}$Sb$_{19}$, we relaxed two of these arrangements with DFT.
The structures differ substantially from the observed ordering, with 34 and 39 In/Sb exchanges among the 184 framework sites of the 216-atom cell \revSI{(Table~S13 of the SI)}.
Sb migrates from 16i to 24k sites and In from 24k to 16i and 6c sites (6c In:Sb occupations of 18:6 and 21:3, instead of 12:12).
As a result, both motifs that define the superstructure are lost: the T/Pn alternation along the 6c chains and the collinear 16i Pn--Pn pairs, which are replaced by Sb--Sb bonds between 16i and 24k sites and between 24k sites.
The number of homopolar bonds per cell increases from 72 In--In and 8 Sb--Sb to 78 and 14, and to 80 and 16, respectively, at the expense of 12 and 16 In--Sb bonds \revSI{(Table~S14 of the SI)}.
DFT places these arrangements 21 and 26~meV/atom above the observed ordering, \textit{i.e.}, 4.4 and 5.6~eV per cell, which corresponds to $0.72\pm0.02$~eV for each In--In plus Sb--Sb pair that replaces two In--Sb bonds.
This bond-counting picture quantitatively supports our hypothesis that the superstructural ordering maximizes heteropolar bonding.
The MEGNet model ranks these arrangements correctly with respect to the observed ordering, consistent with the case of Rb$_8$Ga$_{27}$Sb$_{19}$,\cite{cerasoli_effective_2024} but it underestimates the energy differences by about one order of magnitude (2.0 and 2.6~meV/atom). Therefore, MEGNet-guided searches can generate and rank candidate decorations, but the energy differences between them must be computed with DFT.
Hence, our calculations adopt the lowest-energy decoration found, and none of the searches identified a lower one.
These results point to two distinct classes of decoration changes, with very different implications for configurational effects at synthesis temperatures ($k_BT\approx 70$--80~meV at 800--900~K).
Changes that introduce homopolar bonds cost about 0.7~eV per In--In plus Sb--Sb pair, \textit{i.e.}, about ten times $k_BT$, so disorder of this type should be suppressed, consistent with the exact site occupancy observed in the synthesized A$_8$T$_{27}$Pn$_{19}$ clathrates.\cite{owens-baird_iiiv_2020}
Conversely, exchanges within heteroatomic pairs that preserve the bond counts, as in the alternative $Ia\bar{3}$ decoration of Rb$_8$Ga$_{27}$Sb$_{19}$, cost less than 2~meV/atom over the whole cell, and might therefore be thermally accessible.
Whether this leads to partial disorder on the 6c and 24k sites, and how it affects the free energy of the ordered superstructure, cannot be established from the few configurations computed here. Quantifying configurational entropy would require a dedicated study, \textit{e.g.}, with cluster expansions or machine-learning potentials trained on DFT decoration energies.}

\section{Conclusion}\label{sec13}

This high-throughput analysis predicts {22} stable compounds in the A$_8$T$_{27}$Pn$_{19}$ family \rev{within scalar-relativistic DFT}, of which only three are currently known experimentally. \rev{A few of them lie within 10~meV/atom of the hull, where the classification is marginal, and the Bi-containing candidates are likely destabilized by SOC}.
Trends in the stability with respect to chemical composition reveal that a lower ionization energy in the inclusion species leads to more stable compounds.
\rev{All alkali guests are fully ionized in the clathrate framework; the lower stability of clathrates with lighter guests is therefore not due to incomplete electron donation, but correlates semi-quantitatively with the higher energetic cost of ionizing them. The heavier guests, with the lowest ionization potentials, provide the highest stability. The rattling behavior is instead governed by the guest size: the small Na$^+$ ions move off-center and perturb the framework, whereas Cs$^+$ ions remain confined near the cage centers.}
\rev{Aluminum-containing ATPn clathrates are found to be unstable}, and frameworks containing indium and gallium are most likely to form ternary ATPn clathrates.
Lattice parameter and volume are shown to scale strongly with the atomic numbers of constituent framework atoms and weakly with the atomic numbers of guests.
Superstructural order is carefully analyzed with regard to transformations to each Wyckoff site in the type-I clathrate unit, which achieves fewer pnictide-pnictide bonds in the superstructure by reducing the symmetry group. \rev{Alternative decorations that introduce homopolar bonds cost about 0.7~eV per additional T--T plus Pn--Pn pair, which quantitatively supports the view that the observed ordering maximizes heteropolar bonding, and indicates that the stability of these compounds can only be predicted reliably when the correct decoration is used.}

Synthesis is attempted for compositions containing indium, gallium, arsenic, and bismuth that have not been obtained experimentally to date.
While these attempts did not result in novel clathrates, four distinct crystalline compounds were produced, which are, to our knowledge, previously unreported: Rb$_2$In$_2$As$_3$, Cs$_2$In$_{3.3}$As$_4$, KGa$_{0.4}$Bi$_{1.7}$, and KGa$_{0.3}$Bi$_{0.7}$.
Inability to obtain the target bismuth-containing phases, predicted stable by the standard DFT calculations,  instigated a refinement of the computational framework using fully relativistic SOC, which revealed large changes to formation energies when substantial amounts of bismuth are present in the compound.
{The failure to synthesize other clathrates predicted to be stable in DFT calculations stems from limitations in our theoretical approach, which amount to (i) not searching for unknown competing ternary compounds, (ii) ignoring finite-temperature thermodynamic effects on stability, \rev{including vibrational and configurational entropy,} and (iii) not considering kinetics effects. 
Furthermore, defects, disorder, and coupling to imperfections can add complexity compared to idealized supercell calculations.}
This work provides insight into the ATPn phase space from the perspective of both first-principles prediction and laboratory synthesis. It also highlights the importance of having a tight loop between theory and experiments, as, in some cases, standard high-throughput calculation approaches, commonly adopted in large materials databases,\cite{jain_commentary_2013} fail to predict even the basic ground state properties of certain compounds.

\begin{acknowledgments}
The authors acknowledge support from the U.S. Department of Energy, Office of Basic Energy Sciences, Division of Materials Science and Engineering, Grant No. DE-SC0022288.
\end{acknowledgments}

\section*{Data Availability Statement}
The data that support the findings of this study are openly available in the Materials Cloud Archive, \cite{talirz_materials_2020} record number 2026.XXXX.


\bibliography{bibliography}

\end{document}